\documentclass{si3_osu}
\usepackage{epsfig}
\begin{document}
\def\pb{pb$^{-1}$}
\def\fb{fb$^{-1}$}
\def\ep{e$^+$}
\def\em{e$^-$}
\def\Ohm{${\rm \Omega}$}
\def\mum{${\rm \mu}$m}
\def\muA{${\rm \mu}$A}
\def\musec{${\rm \mu}$sec}
\def\rphi{${\rm r\phi}$}
\def\tilde{$\sim$}
\def\degree{$^o$}

\begin{frontmatter}
\title{The Power Supply System of the CLEO III Silicon Detector}
\author{E.~von~Toerne,}
\author{J.~Burns,}
\author{J.~Duboscq,}
\author{E.~Eckhart,}
\author{K.~Honscheid,}
\author{H.~Kagan,}
\author{R.~Kass$^*$}\corauth{Corresponding author. \\
{\it E-mail address:} kass@ohstpy.mps.ohio-state.edu},
\author{D.~Larsen,}
\author{C.~Rush,}
\author{S.~Smith,}
\author{J.~B.~Thayer}

\address{The Ohio State University, Columbus, OH 43210 USA}
\begin{abstract} 
The CLEO III detector has recently commenced data taking at the 
Cornell electron Storage Ring (CESR).
One important component of this detector is a 4 layer 
double-sided silicon tracker 
with 93\% solid angle coverage. 
This detector ranges in size and number of readout channels 
between the LEP and LHC silicon detectors.
In order to
 reach the detector performance goals 
of signal-to-noise ratios greater than 15:1
low noise front-end electronics
together with highly regulated low noise power supplies
were used.
In this paper we describe
the low-noise power supply system and associated
 monitoring and safety features used by the CLEO III silicon tracker.


\end{abstract}

\end{frontmatter}
\section{Introduction}
The CLEO III experiment\cite{CLEOIII}
is located at the Cornell Electron Storage Ring (CESR), 
a symmetric \ep\em\ collider operating in the 
${\rm \Upsilon(4S)}$ energy range ($\approx$ 10.5 GeV).
It provides access to a rich variety of $B$-meson, tau and charm physics topics.
CLEO III is the third major upgrade of the CLEO experiment in the past twenty
years. This
upgrade had several goals including
 improving the particle identification capabilities to insure
$>3.5\sigma~K/\pi$ separation over  the
entire range of momenta expected from $B$-meson decay,
replacement of 
the charged particle tracking system as most of its elements were more than a
decade old, and a new data acquisition system to accommodate
the expected increase in CESR luminosity and new detector systems.
Several important elements of the CLEO II 
detector were kept for use in CLEO III.
These elements included the CsI calorimeter, 1.5T 
superconducting magnet, and the
muon detector system.

In order to create sufficient space inside the magnet volume for a 
 Ring Image CHerenkov (RICH)
detector
the radius of the charged particle tracking system was reduced from CLEO II's
1 meter to 0.8 meter.  One of the challenges of the CLEO III tracking system
was to provide the same momentum resolution as CLEO II's but use a smaller 
region of space.  To meet this challenge a new silicon detector
and wire drift chamber were designed and fabricated. 
In this paper we first describe the CLEO III silicon detector
and then provide details on the power supply system used to operate it. 

\section{CLEO III Silicon Detector}
\begin{figure}[htb]
\psfig{figure=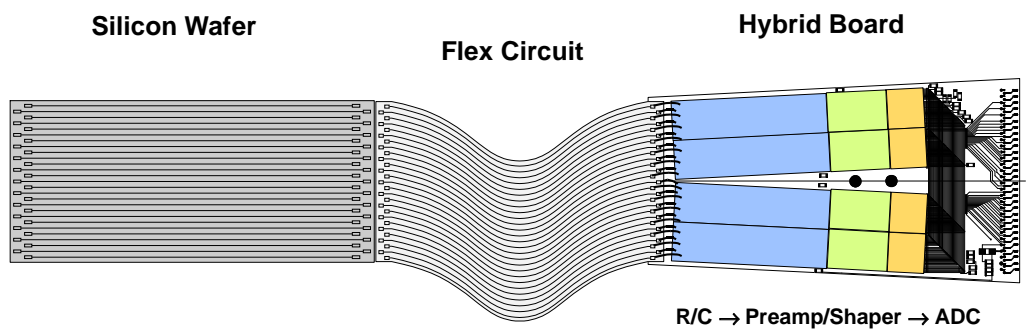,width=5.5in}
\caption{\label{chain}A silicon wafer readout chain.}
\end{figure}

The CLEO III silicon detector has been 
been described in detail in previous publications
 \cite{vertex00,vertex98,IAN,HK}. Here we briefly review its design.
The CLEO III silicon detector consists of 4 layers of
double-sided silicon microstrip sensors that 
form barrels around the beampipe. A total of 447 identical sensors 
are daisy-chained into 61 ladders that are
read out on both ends. Each of the 122 readout chains consists of sensors, 
a pair of flex cables, a hybrid board that hosts resistor/capacitor (RC), 
frontend amplifier (FE) and 
backend digitizer (BE) chips (Fig.~\ref{chain}), a 
portcard, and subsequent VME readout and control boards.
\vspace*{1.8cm}

The sensors are DC coupled; bias resistors and AC coupling capacitors
are located on separate 
chips (RC). Each sensor contains 511 $r-\phi$ strips and
511 $z$ strips. Thus there are 125$\times 10^3$ readout channels in
the CLEO III silicon detector.
 Three voltages, Nstrip, Pstrip,
and Pstop, are necessary to bias the 
sensors. The  Nstrip and Pstrip voltages deplete the sensor bulk.
The Pstop voltage biases ``atolls'' around the strips on the n-side and thus
reduces the capacitance and supports the depletion process. The sensors
are oriented such that
the p-strips measure the $z$ coordinate (parallel to the beam axis)
while the n-strips measure the $r-\phi$ coordinate.

To minimize tracking uncertainties due to multiple scattering
the mass of the detector was made as low as possible 
(radiation length $<$ 1.6\%). 
To achieve this end,
the detector readout electronics are not in the tracking volume but at the
ends of the ladders.  Thus most of
the radiation length is due to the silicon sensors. 
In order to achieve good tracking efficiency a signal to noise 
ratio of $>15:1$ in all layers is desired.
However, due to the considerable length of the sensor ladders, 
up to five sensors are ganged together, 
low noise front-end electronics and low detector capacitance are required
to achieve this signal-to-noise goal.
The frontend electronics chips are located on hybrid boards that are
mounted outside the active tracking volume on the mechanical support cones.
 Two flexible (``flex'') cables are used to connect n and p-side readout strips
of the sensor to the hybrid. The flex design uses two metal layers
patterned with 
traces on both sides of a 25~$\mu m$ kapton substrate~\cite{flex}. 

 BeO was chosen to be the hybrid substrate due to its excellent
thermal and mechanical properties. Each hybrid has four  
 RC, FE, and BE chips on the top and bottom surfaces as well as
bypass capacitors, resistors for the chip voltages, a temperature
monitor circuit, and two connectors for power and data lines.
Each RC chip hosts 128 bias resistors and AC coupling capacitors on a
500 \mum\ thick quartz substrate.
Due to the common ground for n and p-side electronics the voltage
across the coupling capacitor 
is about half the depletion voltage. This is well below the measured
RC breakdown voltage which is in the range of 75 - 100 V.
The FE chip consists of a preamplifier, a CR-RC shaper, and a
variable gain stage. A 5V supply (AVDD) and a 2.5V supply 
(AVDD2) are used to operate the FE chip. 
Amplified signals from the FE chips are
transferred to the BE chips for digitization. The BE chip is 
a modified version of the  LBNL SVX2b
 design\cite{svx}. Each BE chip hosts 128 8-bit
Wilkinson ADC's, comparators and a First-In-First-Out (FIFO) buffer. 
The comparators allow for sparsified readout. 
The BE chip is powered by a 5V supply (DVDD).
Most of the power consumed by the silicon detector, up to 4 Watts per
hybrid, is generated by the electronics
on the hybrids. The heat generated on the hybrid
is transferred to the support cones
through a copper post on the hybrid.

Signals and supply voltages 
are routed to the hybrid through intermediate boards called portcards.
The portcards are located in eight crescent-shaped
crates that form  full rings around the beampipe. These cards
are located about 1 meter
away from the interaction region.  
Each crescent crate contains up to 16 portcards and one buffer
card that serves as a signal repeater for CLEO III slow control signals.
The portcards host the ADC's and DAC's used by the slow control
system. These cards are powered by a 5V supply  (V5PC) and 
a 6V supply (V6PC). The 6V supply is used to generate low current voltages that supplement
the FE and BE chips supply voltages. 
The readout chains terminate in VME based data boards equipped with 
a BE chip readout sequencer,
data FIFOs and a digital signal processor for formatting the data.
Each data board services four readout chains.

The slow control process resides in a MVME 2432 CPU \cite{MVME} inside the master slow control crate.
This crate also contains the interface boards
necessary to control the portcards. Two VME repeater boards located in the
same crate communicate with corresponding repeater boards in the two power supply crates. 
This setup with only one slow control process residing in one crate
CPU makes powering the silicon detector almost independent of 
computer network problems and simplifies the slow control program structure. 
\begin{table}
\caption{\label{tab_voltagetypes}Voltages necessary to operate the silicon
detector and power supply cards. The currents in this table are typical
half-ladder values.}
\vspace*{1ex}
\begin{tabular}{c c c c c}\hline
Voltage &  Description & Voltage    & Total Current    &            \\
Name    &              &  (Volts)  & (Amps)   & Control \\ \hline
Nstrip  & Sensor n-side bias & 50 & \tilde$10^{-6}$\    & DAC  \\
Pstop   & n-side barrier & 35 & \tilde$10^{-6}$\    & DAC  \\
Pstrip  & Sensor p-side bias & -50 & \tilde$10^{-6}$\    & DAC  \\
V5PC    & Portcard Supply &  5.0 & 0.15     & Switch  \\
V6PC    & BE/FE currents &  6.0 & 0.1     & Switch  \\
AVDD    & FE chip supply &  5.0 & 0.25     & Switch   \\
AVDD2   & FE chip preAmp+shaper &  2.5 & 0.1-0.8  & Switch \\
DVDD    & BE chip supply &  5.0 & 0.1      & Switch  \\
V5PISO  & PS card analog &  5.0 & --        & --  \\
V12+    & PS card analog &  12.0 & --        & --  \\
V12-    & PS card analog & -12.0 & --        & --  \\
VME +5  & PS card VME    &  5.0  & 0.2  & monitor card \\
\hline
\end{tabular}
\end{table}

\section{CLEO III Silicon Power Supply System}
\subsection{Silicon Power Supply System Overview}
\begin{figure}[htb]
\psfig{figure=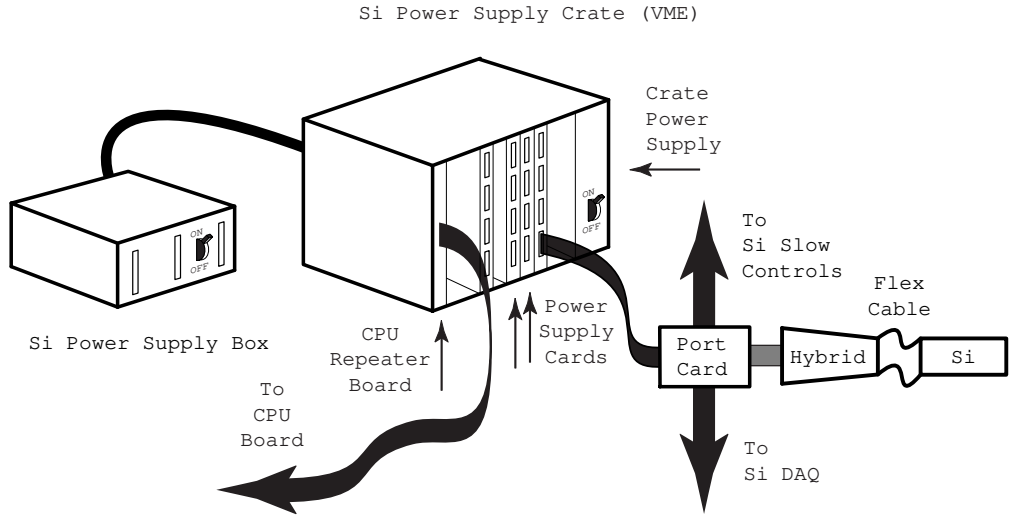,width=5.5in}
\caption{\label{blockd}The principal components of the 
CLEO III silicon power supply system.}
\end{figure}

The CLEO III power supply system provides the voltages
to each of the above subsystems in an integrated fashion.
The CLEO III power supply system
consists of two modified VME crates,
six power supply boxes, thirty four power supply cards and software.
The power supply cards, boxes and the J3 VME backplanes were custom 
designed to meet the needs of the CLEO III silicon project. In addition
to the specialized hardware, software has been developed specific for the
 power supply cards and boxes and for communications between
the power supply system and the rest of the CLEO III experiment.
A diagram with the principal power supply system components
is given in Fig.~\ref{blockd}
In the sections below we describe each of these components in detail.

\subsection{Power Supply Boxes}
A power supply box contains the DC power supplies necessary to operate the
 the portcard, the hybrid, and the analog section of the power
supply card as well as those necessary to bias the detector
and monitor the status of the supplies.
Linear power supplies\cite{ACOPIAN} chosen for their excellent noise
performance are used to power the silicon detector.
There are 9 supplies inside each box. 
This relatively large number of supplies results from our decision to use
separate supplies for digital and analog sections of the hybrid
and portcard.  This prevents noise coupling through
common power leads. All supplies are floating at the supply end, ground
is at the detector.
All supplies are current
limited and over-voltage protected.
Each box can supply up to 28 hybrids (7 power supply cards). The
entire silicon detector is powered by 6 boxes, three on the east and
three on the west side of the detector. One box on each side services
hybrids in layers 1 and 2, one box layer 3 hybrids and one box layer 4
hybrids.

A power supply box can be controlled either through an internal
 monitor card,
driven by an RS485 serial line, or by front panel switches. 
The monitor card also records 
the temperature of the supplies and voltage levels
at the power supply box output. In remote mode these voltages 
and temperatures
are compared with programmable upper and lower limits and if a supply is
out of bounds a relay is activated that shuts the supply off. 

Each box can consume up to 200 Watts. Much of the power is transformed into
heat inside the box. Since the power supply box is a closed system the
heat can not be taken out by circulating air alone.   
The heat is taken out via copper plates and water cooling lines on
the top and bottom of the box. 
Two fans inside the box increase air circulation and improve the cooling of
the power supplies. Studies using 15.6\degree C water 
flowing at 1.5 liters/min through the
bottom and top copper plates show that the ambient temperature inside
the box stays well below 50\degree C at full load. 

The boxes are located about 2 meters away from the power supply
crates. Each power supply box is connected to the back of a power supply crate 
by 36 cables.
All cables are attached to the back of the power supply box. 
Of the 36 cables 20 are current carrying and 16 are for remote sensing.
All voltages except the power supply card supply voltages (5V, +/- 12V) are
remote sensed at the back of the power supply crate.
A power supply box weighs about 150 pounds (68 kg) and is operated
with 208 V AC. 

Before installation into
the CLEO III experiment each power supply box was required to
operate continuously at 120\% load for at least three days. 
No power supply failures or cooling related
problems have been observed during the first six months of 
operation of the silicon detector.

\subsection{Power Supply Crates}
There are two power supply crates in the CLEO III silicon detector system.
These are 9U VME crates with a custom J3 backplane.
Cables from the back of the power supply boxes
terminate in connectors that plug into a custom-made J3 backplane.
This backplane has a row of J3 connectors attached on the inside of
the crate. Each VME crate contains 17 power supply cards and a VME 
repeater card. 
The repeater card
and VME section of the power supply cards
 are powered by 5V, high current, high frequency switching power 
supplies\cite{VICOR} that are located inside the VME crates.
This voltage is controlled by a monitor board that allows remote
switching of the VME 5 V. The monitor board also 
interlocks the crate power to
the cooling system of the crates and of the silicon detector.
The VME crates are water-cooled. 
The system is designed such that all power transfer to the detector 
is shutdown in a controlled fashion
when a cooling interlock is broken or the crates
are turned off.

\subsection{Power Supply Cards}
The power supply cards distribute the voltages
necessary to operate the silicon detector in a fashion consistent
with its low noise goals. For safety reasons each card also monitors,
using a single 12 bit ADC,
the relevant voltages, currents, and temperatures. Should a voltage,
current, or temperature exceed its onboard programmable limit (upper
or lower)
the card will automatically trip the offending supply 
circuit and send a message
to the CLEO III control system.
The power supply cards reside in the power supply crates. There are 34 cards
in the entire system, with 17 in each crate. 
Each card is divided into a digital VME part and an analog
part. The analog part is
subdivided into four identical sections with each section servicing
one detector readout chain.
Power comes in through the J3 VME connector and is distributed to the
four sections. Each section is connected to a portcard, approximately
twenty five feet away, via a 50 conductor
flat cable~\cite{beldon}. In order to minimize voltage losses due to
the resistance of the cable conductors a fifty foot length of cable
is folded in half and the portcard connector is put in the middle of
the cable and power supply card connectors on each of the cable ends.

\subsubsection{VME and control section of power supply card}
The power supply cards communicate with the rest of the experiment
using the VME protocol.
The VME section of each card consists of VME buffer chips, an
ALTERA 10K-30 chip (onboard state machine), 
and a 30MHz oscillator to clock the state machine.
The state machine controls ADC and DAC functions and  performs on-board
functions through VME commands. An independent state machine
 process performs
voltage, current and temperature monitoring tasks. 
The state machine program is described in the next section. 
The VME section is powered by a dedicated  5 V supply located in 
the power supply crate.

Because this supply is a switching supply, it is isolated
from the voltages 
 that go out to the hybrids in order not to compromise the
noise performance of the detector.
The board state machine exchanges signals with the 
analog section on the card through electrically isolated optocoupler chips.
This electrically isolates the VME section from the rest of the board. 
The DAC settings and ADC values are transferred on a 12 bit bi-directional
data bus.  The control and status signals are unidirectional.
 
There are 32 switches, 12 ADC lines, 8 DAC lines, and 10 select
lines  on a power supply card. In order to save space on the card 
 a design with two address busses (CSEL(0:3), SEL(0:5)) and 
programmable decoder chips~\cite{XLINX} was chosen. 
Each switch or DAC-select line is identified by a unique
set of address words. 
The address values are decoded by four programmable decoder chips
that provide the switch-select and DAC-select signals for one
section.

\begin{figure}
\psfig{figure=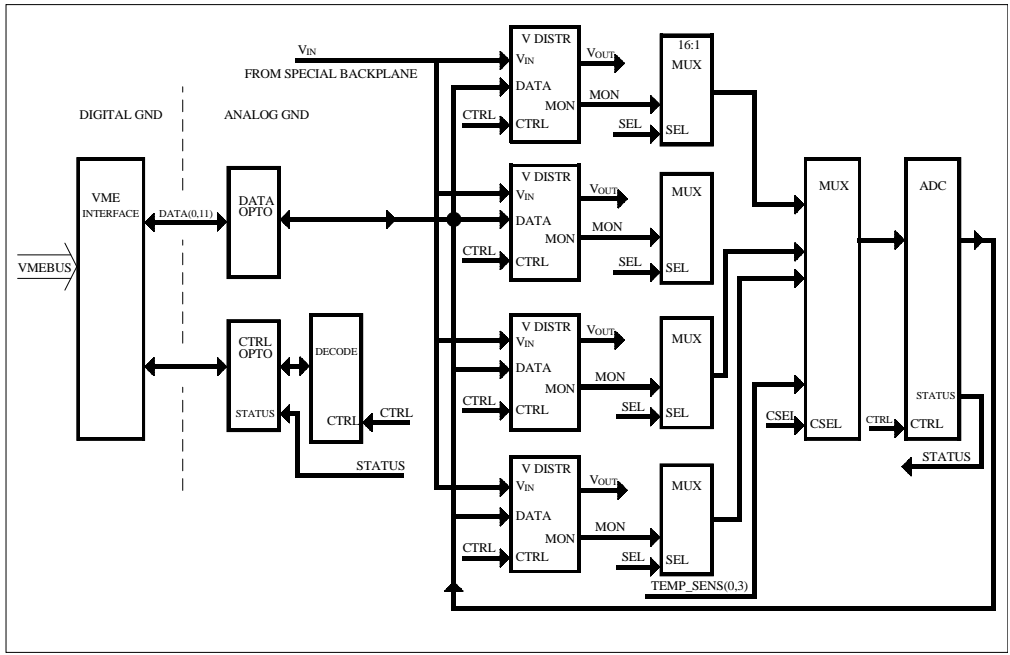,width=5.25in}
\caption{\label{psboard_diag}Schematic view of the power supply 
card structure.}
\end{figure}

Signals are routed from the analog
section of the board to the ADC via 
a series of multiplexers.The multiplexers are controlled by the same address
lines as the DAC's and decoder chips.
The address space has been divided between multiplexer and
decoder functions so that the multiplexer can be operated
without activating the decoder chips and vice versa.
A schematic view of this system is given in Fig.~\ref{psboard_diag}.

\subsubsection{Analog section of power supply card}
\begin{figure}[htb]
\begin{center}
\psfig{figure=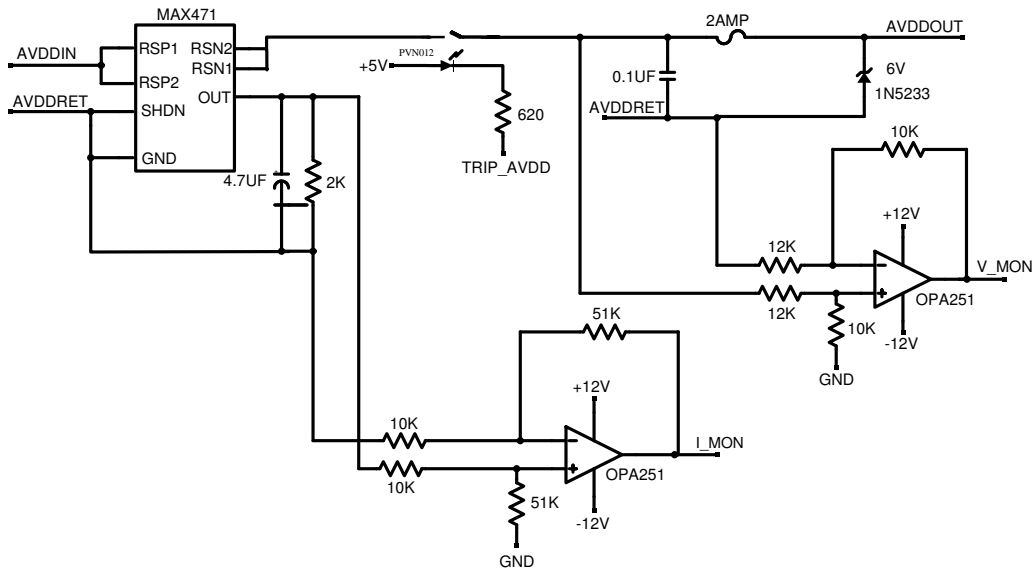,width=5.5in}
\caption{\label{Avdd}Schematic of the AVDD (5 V) circuit.}
\end{center}
\end{figure}

Each card contains four identical analog sections. 
The analog sections of the board distribute the voltages
 to the portcards, hybrids and sensors, host the 
on-off switching network for the 
 portcards (V5PC, V6PC) and hybrids (AVDD, DVDD, AVDD2), the bias voltage
(Nstrip, Pstrip, Pstop) circuits, and hybrid temperature monitoring
circuits. In Fig.~\ref{Avdd} the circuit used for the AVDD, the
analog +5V supply, is shown. 

The MAX471, shown in Fig.~\ref{Avdd} is a high side current monitor.
It mirrors the high side to a ground referenced
current.  It introduces a series resistance of 0.07 \Ohm.
The PVN012 is a solid state switch with an on resistance of
0.04 
\Ohm. The  voltage difference between the power lead and the
return line is monitored. Both the current and the resistance
of the power leads are known, thus the voltage at the load can be calculated.
The result is a software remote sense.
 This circuit is representative of the 
ones used for DVDD, AVDD2, V5PC, and V6PC. 

\begin{figure}[htb]
\psfig{figure=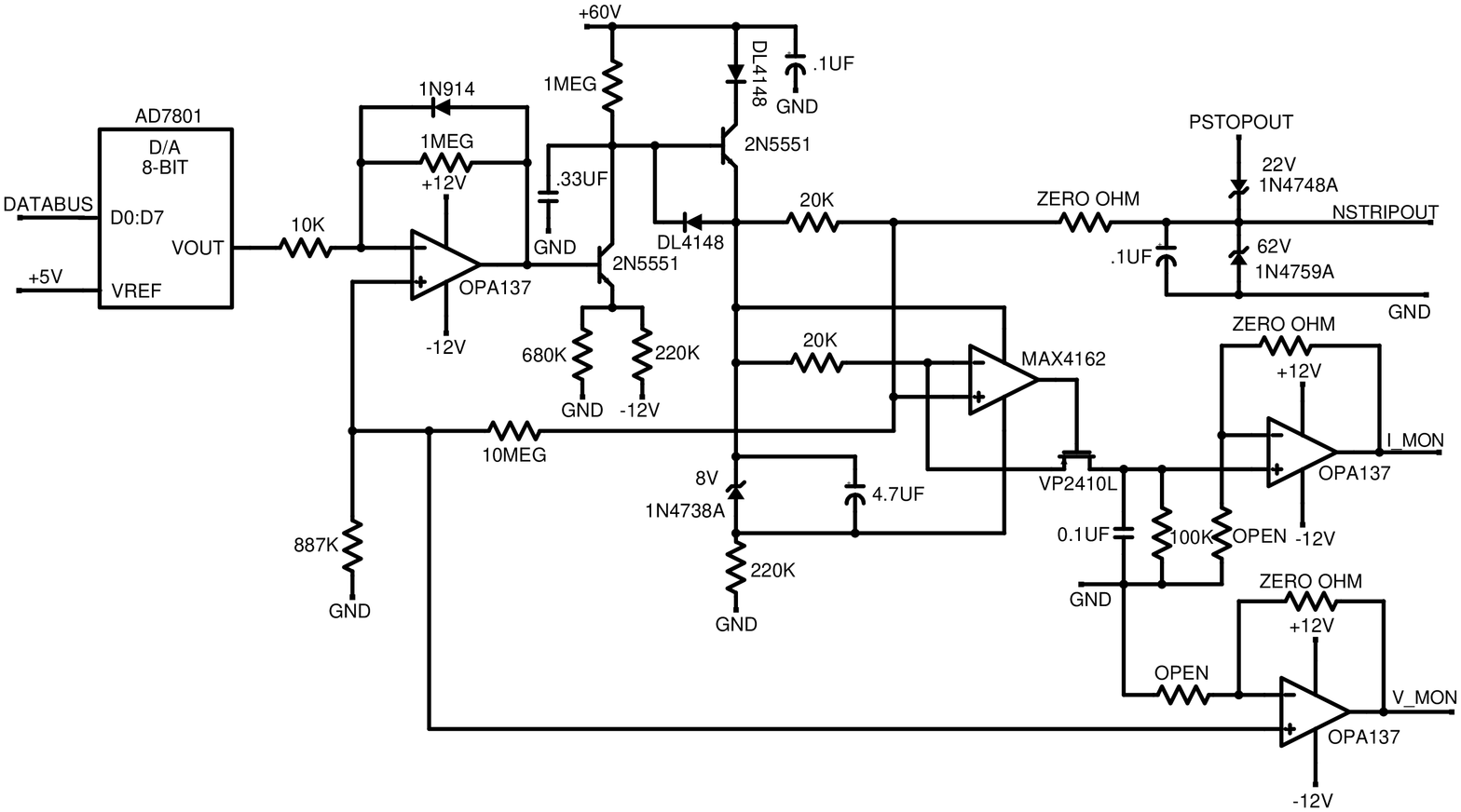,width=4.2in,angle=90}
\caption{\label{nstrip}Schematic of the Nstrip bias circuit.}
\end{figure}

\begin{figure}[htb]
\psfig{figure=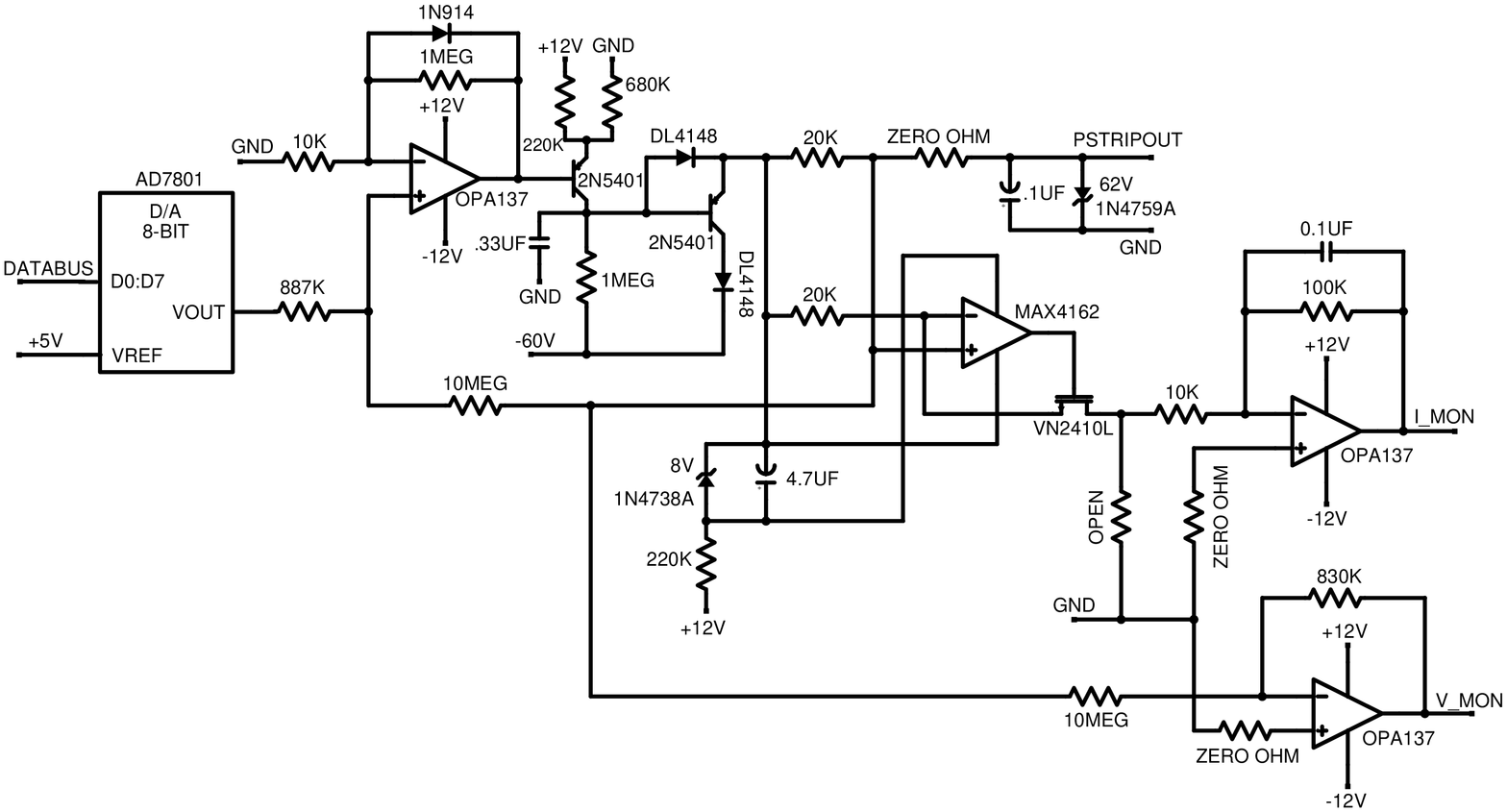,width=4.2in,angle=90}
\caption{\label{pstrip}Schematic of the Pstrip bias circuit.}
\end{figure}

\begin{figure}[htb]
\psfig{figure=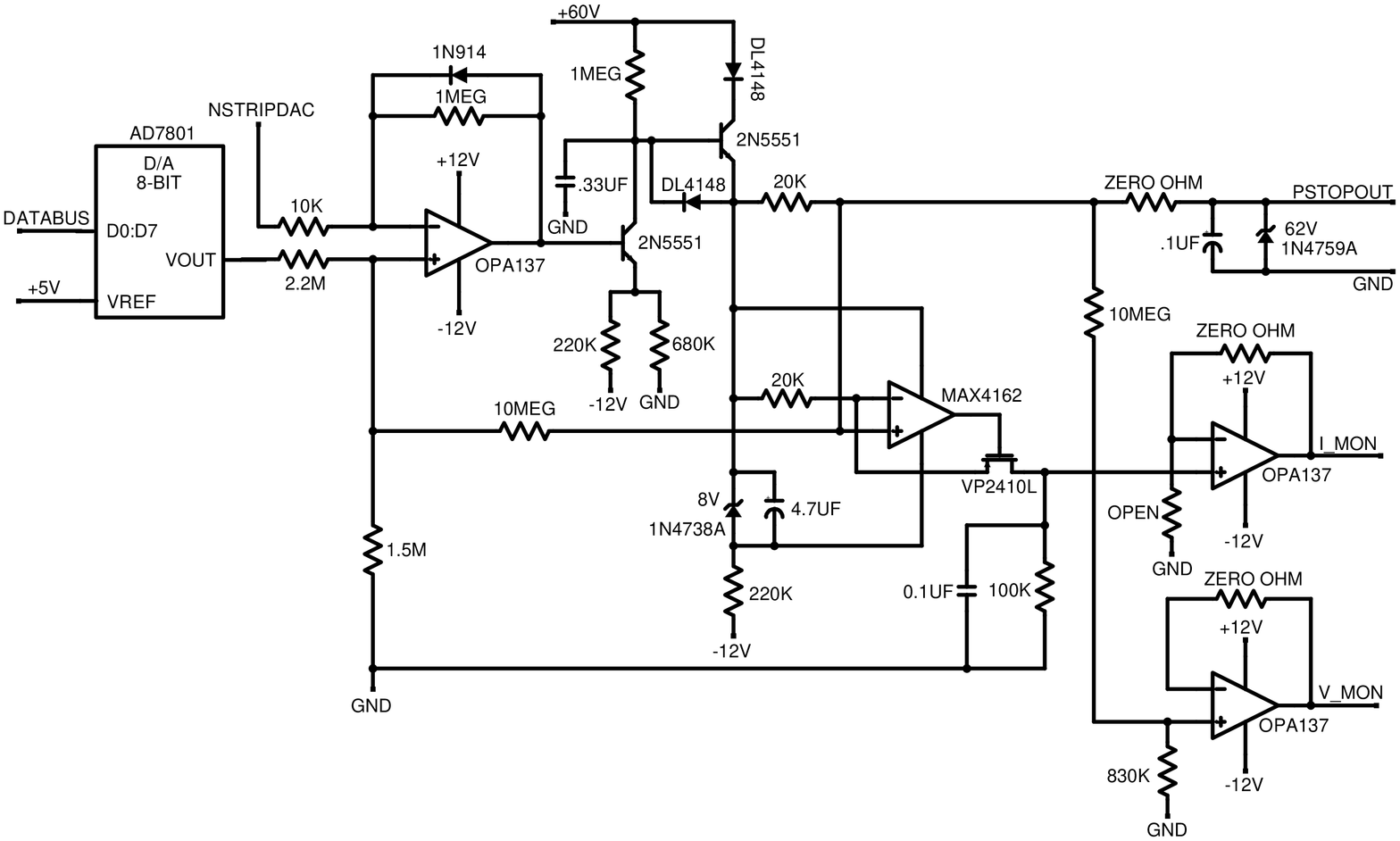,width=4.2in,angle=90}
\caption{\label{pstop}Schematic of the Pstop bias circuit.}
\end{figure}

The three bias voltages (Nstrip, Pstrip, Pstop) are generated 
on the power supply
board from the 60 V power supplies located
in the power supply boxes. The schematics for these circuits
are shown in Figs.~\ref{nstrip},~\ref{pstrip}, and~\ref{pstop}
respectively.
The typical operating values
for these voltages are
given in Table \ref{tab_voltagetypes}. The circuit designs for
the three voltages are similar. The output voltage is controlled
by a DAC; voltages and currents are measured by similar circuits.
The bias voltage can be ramped in steps of \tilde 0.2 V.
Sudden large bias voltage changes from ($e.g.$) a power outage
can in principle damage the RC chips.
To protect against this situation the bias circuits
provide a 2 second settling time without adding a large series resistance
to the output. 

The analog sections that read the current and
voltage levels have been designed to 
conform to the low noise requirements of the silicon detector.
Each of these sections transforms and/or buffers the analog input into
a voltage that can be digitized by the ADC. 
A total of 68 values are read by the board and transferred to the ADC
through the two analog multiplexer (MUX) levels. 
Each analog section has one 16-input MUX
for 8 voltage and 8 current channels. 
The section multiplexer passes  signals on to the top-level MUX.
The top level MUX  selects the ADC input from the four section
multiplexer outputs. The hybrid temperature channels do not fit into
the section multiplexer and are directly connected to the
top-level MUX.

All hybrid and portcard voltage lines
 are protected by  fuses on the power supply card. 
The fuse  ratings are limited by a
second set of fuses on the portcards, which are less accessible.
Therefore fuse ratings for the
power supply card fuses were carefully chosen so that these fuses blow
before the portcard fuses.

\subsection{Power Supply Card Software}
The power supply card state machine is programmed in AHDL~\cite{AHDL}
to perform  the three main tasks described below.
A general 32 bit register controls most of the features to be described, e.g.
the timeout constant and the speed of the monitor process.

1.) Communicate with the CLEO III experiment via VME. 
The state machine interprets VME commands which are sent by
the master slow control CPU.
All DAC's, ADC's and switches can be operated 
remotely through these VME commands. 
The ADC readout has been made stable against short-time fluctuations
by using the median value of five consecutive digitizations of an
individual channel as the value returned to the slow control system.

2.) The program is equipped with a counter that measures the time
of the last interaction with the slow control system. Each VME interaction resets
the counter. If the counter reaches an externally adjustable value 
the state machine turns off all voltages controlled by this card.
In current operation the timeout occurs after 2 minutes. 
The silicon detector has benefitted several times from this feature.

3.) A major part of the program
is an independent readout loop that
 monitors
all channels by digitizing the voltage, current and temperature
values and
comparing them to limits that
are stored in internal registers. 
The comparison against lower limit values is also an option.
In order to minimize readout
fluctuations each channel is read out 4 times in a row and a
limit is considered surpassed if 3 out of 4 readouts are above limit. 
During
each readout loop the channels are compared against warning and trip
levels.
If a warning limit is surpassed a flag is set in an internal register and, if
permitted by the slow control system, an interrupt is set.
If a trip limit is surpassed the section is turned off and
the address of the channel that caused the trip is stored in an internal 
register. 

All limit registers are initialized automatically when the power supply
cards are powered on. Each limit value can be updated externally.
A readout loop consists of 272 digitizations and ADC readouts and takes
a minimum of 3 msec.  Most of this time is consumed by the digitization
process.
At this speed a small noise excess of the order of 100 \em\ ENC has been 
observed in the hybrid frontend electronics of the silicon detector.
The speed of the monitor process can be slowed down externally 
in order to remove this noise which is caused by multiplexer switching.
For readout loops longer than 500 msec no additional noise is 
introduced into the system. During normal operations the monitoring is 
performed with 1 sec long readout loops. 
In addition each channel is readout by slow control about every ten seconds.

A power supply card can also turn off the AVDD, AVDD2, DVDD, V5PC, V6PC,
and bias supplies located in its corresponding power supply box. 
A buffered signal from the card's state machine
is brought from the front of the power supply card to the power supply
box. This signal is daisy chained so that only one cable runs between
the power supply cards and a power supply box.

\subsection{Performance}
The silicon detector has been taking physics quality data since July 2000
accumulating approximately 4.5 \fb\ of data to date.
The silicon power supply system has proven to be stable in daily
operations. Failures of system components have been almost exclusively
restricted to fuses.

The frontend electronics has achieved its design goals.
The single channel noise by layer ranges between 400 and 800
electrons. The z-side shows slightly lower noise than the
$r\phi$-side due to the  different sensor strip capacitances for n-side
(=$r\phi$) and p-side (=z).
The common-mode noise is of the order of 400 \em\ ENC.
The noise values were obtained using the standard conversion of ADC
counts to electrons which is known to about 20\%.
 
Signal-to-noise ratios, which are independent of this calibration,
were obtained from a high-statistics Bhabha sample recorded in summer
2000.  Neighboring groups of channels above pedestal were merged into
clusters.
The sum of the pedestal-subtracted signals over all strips in a cluster
was taken as the  pulse height. Only clusters with 3 strips or less were
considered. Tracks were restricted to a region of $|\cos\theta| < 0.174$
in order to select tracks with almost perpendicular incidence angle.
The pulse height of silicon clusters associated to these tracks were
divided by the channel noise averaged over all strips in the
cluster.  Loosening the cluster width or polar angle requirements
increases the signal-to-noise ratios.
Noise values and signal-to-noise (S:N) ratios for all 4 layers are
given in Table \ref{sn}.
\begin{table}
\caption{\label{sn}Common mode subtracted signal-to-noise ratios and
noise levels by layer and by readout side. The most probable value of
the signal-to-noise distribution is quoted as the ratio.}
\vspace*{1ex}
\begin{center}
\begin{tabular}{c  c c  c c}  \hline
Layer & \multicolumn{2}{c}{signal-to-noise}  &
\multicolumn{2}{c}{noise (100 \em)}\\
      & r-$\phi$  & z      & r-$\phi$  & z      \\
\hline
1 & 27.9 & 34.4 & 6.1 & 6.1 \\
2 & 29.9 & 37.4 & 5.9 & 4.4 \\
3 & 19.4 & 27.3 & 8.2 & 5.7 \\
4 & 20.1 & 22.6 & 8.2 & 6.5 \\
\hline
\end{tabular}
\end{center}
\end{table}

\section{Summary and Conclusions}
The design and status of the power supply system of the CLEO III silicon
detector has been presented in detail here. 
The silicon detector has been taking physics
quality data since July 2000. The power supply system has been stable in
daily operations.  The noise performance of the silicon detector and
its power supply system has been very good with 
400-800 \em\ ENC common-mode subtracted noise.
The frontend electronics has reached its signal-to-noise goals of better than
15:1 in all layers. 
\section{Acknowledgments}
We would like to thank Vijay Sehgal for the excellent electronics 
shop support, Bob Wells for his help with the mechanical aspects of this
project, and The Ohio State University's physics department machine shop.    
This work was supported by the 
Alexander-von-Humboldt Stiftung, The Ohio State University,
 and the U. S. Department of Energy.

\vfill
\eject
\end{document}